  \documentclass[epjST]{svjour_MODIFICADO}
  \usepackage{url}
  \usepackage[english]{babel}
  \usepackage{microtype}
   \DisableLigatures[f]{encoding = *}
 \usepackage[usenames]{color}
 \definecolor{dark-green}{RGB}{0,100,0}
 \definecolor{dark-red}{RGB}{139,0,0}
 \definecolor{dark-blue}{RGB}{0,0,139}
 \usepackage{amstext, amssymb, amsfonts, amsmath}
 \usepackage{bm}        
 \usepackage[mathscr]{eucal}
 \usepackage{exscale}   
 \usepackage[dvips]{graphicx}
 \usepackage[dvips, verbose, breaklinks, hyperindex]{hyperref}
 \hypersetup
   {
    bookmarks  = true,     
    colorlinks = true,     %
    urlcolor   = blue,     %
    filecolor  = cyan,     
    linkcolor  = blue,     
    citecolor  = blue,     
    unicode    = true,     
    linktoc    = all,      
    hyperfigures = true,   
   }
\hypersetup
{
  pdfauthor   = {Leonardo Jos\'e Lessa Cirto},
  pdfsubject  = {Statistical Mechanics, Nonextensive Statistical Mechanics},
  pdfkeywords = {HMF, HMF-alpha-XY, Classical Heisenberg Model, Statistical Mechanics, Nonextensive Statistical Mechanics, Molecular Dynamics Simulation},
  pdftitle    = {Teoria},
}
  \usepackage[hyphenbreaks]{breakurl}
 \usepackage{cite}
    \usepackage{anysize}
    \marginsize{1.5cm}{1.5cm}{1.0cm}{0.50cm}
 \usepackage{times}
 
 \usepackage{caption}
 \newcommand{\de}[1]{\left(#1\right)}
 \newcommand{\comu}[1]{\left[#1\right]}
 \newcommand{\Ntil}{\widetilde{N}}
 \newcommand{\rd}{\mathrm{d}}
 \newcommand{\media}[1]{\left\langle #1 \right\rangle} 
 \newcommand{\Espaco}{\rule[-5mm]{0mm}{13mm}}
  \graphicspath{ {./FIG/} }
\begin{document}
\title{Thermodynamics is more powerful than the role to it reserved by Boltzmann-Gibbs statistical mechanics}
\author{
Constantino Tsallis\inst{1,}\inst{2}\fnmsep\thanks{\email{tsallis@cbpf.br}}
\and
Leonardo J.L. Cirto\inst{1}\fnmsep\thanks{\email{cirto@cbpf.br}} }
\institute{
Centro Brasileiro de Pesquisas Fisicas (CBPF) and National Institute of Science and Technology for Complex Systems (INCT-SC), Rua Dr.~Xavier Sigaud 150, 22290-180 Rio de Janeiro-RJ, Brazil
\and
Santa Fe Institute, 1399 Hyde Park Road, Santa Fe, NM 87501, USA}
\abstract{
We brief{}ly review the connection between statistical mechanics and thermodynamics.
We show that, in order to satisfy thermodynamics and its Legendre transformation mathematical frame, the celebrated Boltzmann-Gibbs~(BG) statistical mechanics is suff{}icient but not necessary.
Indeed, the $N\to\infty$ limit of statistical mechanics is expected to be consistent with thermodynamics.
For systems whose elements are generically independent or quasi-independent in the sense of the theory of probabilities, it is well known that the BG theory (based on the additive BG entropy) does satisfy this expectation.
However, in complete analogy, other thermostatistical theories (\emph{e.g.}, $q$-statistics), based on nonadditive entropic functionals, also satisfy the very same expectation.
We illustrate this standpoint with systems whose elements are strongly correlated in a specific manner, such that they escape the BG realm.
}
\maketitle
\section{Introduction} 
As an enshrined scientist, Einstein, in 1949,
expressed his appreciation of classical thermodynamics, thus 
sharing the deep impression this theory has had upon him.
In his words~\cite{Einstein1949,KleinScience1967}:

\emph
{
A theory is the more impressive the greater the
simplicity of its premises is, the more different
kinds of things it relates, and the more extended
is its area of applicability.
Therefore the deep impression that classical thermodynamics made
upon me. It is the only physical theory of universal
content concerning which I am convinced that,
within the framework of applicability of its basic
concepts, it will never be overthrown.
}

Thermodynamics is the theory of everyday phenomena. 
Many of its variables (volume, pressure, temperature, viscosity) and a large part of its applications (refrigerator, steam engine, batteries)
are known by both scientists and nonscientists.  
Although we usually speak about thermodynamical \emph{laws}, thermodynamic itself is not
a set of fundamental \emph{Laws of Nature} in exactly the same sense that Newton's law and Maxwell's equations are.
We refer to the fact that thermodynamics is consistent with all such laws and, in some sense, covers them all~\cite{Eddington1929}.
We may say that it is the theory which is more widely connected with all fundamental physical laws, 
and it finds its way into many scientific fields, from elementary particles to large scale astrophysics.
The main concern of classical thermodynamic is the relationship between macroscopic variables, 
as, for instance, the Boyle-Mariotte law~$P \propto 1/V$.
Today it is understood that the form of its relations
is consistent with the underling microscopic laws governing the constituents of the system.
It is in many ways universal, and crucial aspects of it are, remarkably enough, valid regardless the particular model.
In between the level of the microscopic description of a physical system and the level of its thermodynamical macroscopic
relations there is statistical mechanics.

The goal of statistical mechanics is, starting from 
the microscopic natural rules (classical, relativistic, quantum mechanics, chromodynamics) and adequately using
probability theory, to arrive to the thermodynamical relations. Along these connections between the macro- and micro- worlds, the ultimate link is made through the fundamental concept of~\emph{entropy}.
This finding, accomplished against a stream of criticism, surely is one of
the most powerful and fruitful breakthroughs of the history of physical sciences. It was achieved by Boltzmann in the last three decades of the nineteenth century.
His result, currently known by every pure and applied scientist,  
and carved on his tombstone in Vienna, namely,
\begin{equation}
S_{BG} = k_B\ln W
\label{eq:Tombstone}
\end{equation}
is the mathematical link between the microscopically fine description (represented by~$W$, the total number of accessible microscopic states of the system)
and the macroscopic measurable quantities (represented by the entropy $S_{BG}$, the very same quantity introduced by Clausius in order to complete thermodynamics!).
Equation~\eqref{eq:Tombstone} has been explicitly stated in this form for the first time by Planck, 
but was clearly known by Boltzmann.
The index~$G$ stands for Gibbs, who put Boltzmann's ideas forward 
and overspread the (classical) statistical mechanics concepts through his seminal book~\cite{Gibbs1902}.
Equation~\eqref{eq:Tombstone} is a particular instance of a more general one, namely
\begin{equation}
S_{BG} = - k_B\sum_{i=1}^{W} p_i \ln p_i
\label{eq:SBG}
\end{equation}
When every microstate is equally probable, \emph{i.e.}, when $p_i=1/W\,\, \forall\, i$, we recover Eq. \eqref{eq:Tombstone}.
Evidently quantum mechanics was unknown to Boltzmann and it was just birthing when Gibbs' book was published.
It was left to von Neumann to extend Eq.~\eqref{eq:SBG} 
in order to encompass quantum systems. 
He showed that the entropy for a quantum system should be expressed 
by using the density matrix operator~$\widehat{\rho}$, namely
\begin{equation}
S_{BG} = -k_B \mathrm{Tr}\comu{\widehat{\rho}\,\ln\widehat{\rho}\,}
\label{eq:Entropia_von_Neumann}
\end{equation}
sometimes referred to as the Boltzmann-Gibbs-von Neumann entropy.
Notice indeed that the above equation recovers Eq. \eqref{eq:SBG} when $\widehat{\rho}$\, is diagonal.

The optimization of the entropy with appropriate constraints provides the thermal equilibrium distribution, namely the BG exponential distribution, whose consequences are consistent with classical thermodynamics.
In what follows we shall, however, see that entropic functionals different from the BG one must be used in order to satisfy thermodynamics for complex systems which violate the probabilistic independence (or quasi-independence)
hypothesis on which the BG entropy is generically based. This is typically the case whenever there is breakdown of ergodicity.
\section{More about entropy}
\label{Sec:Entropies}
The fundamental bridge between the macroscopic thermodynamical variables
and the microscopic world 
is the entropy.
Within statistical mechanics the entropy is a \emph{functional} (of the probabilities) whereas
within classical thermodynamics, as originally imagined by Clausius, it is the state function demanded by the Second Law.
However, after Shannon's insight within the theory of communications, entropy is no longer a concept exclusively related to classical thermodynamics.
In some sense we may say that the entropic \emph{functional} has its own life.
Nevertheless, when dealing with the entropy as the bridge linking the microscopic
and the macroscopic worlds, there are constraints that bind the functional to be used as the physically appropriate entropy.
Herein we focus on the mathematical expression of the entropy by imposing the 
constraint that it must be an \emph{extensive} quantity, \emph{i.e.}, proportional 
to the system size~$N$. Why should this be so as a thermodynamical requirement will be addressed below, in Section~\ref{Sec:Thermodynamic_Entropy}.

It is straightforward to verify, using  Eq.~\eqref{eq:Tombstone}, the extensivity of the entropy when a physical system
belongs to the so called \emph{exponential class}, meaning by this those systems
whose number of admissible microstates increases exponentially with $N$, like 
$W\de{N}\sim \mu^{N}\,\de{\mu>1}$ in the $N\to\infty$ limit.
Those systems generically exhibit weak correlations between their elements, 
including, as a limiting case, the probabilistically \emph{independent} systems, those with no correlations at all (\emph{e.g.}, a classical ideal gas, or a set of noninteracting spins).
Moreover, it is algebraically very simple to verify that~$S_{BG}$ 
is not only extensive for systems of the exponential class, but also \emph{additive}, according to Penrose's definition~\cite{Penrose1970}.
Indeed, if $A$ and $B$ are two probabilistically independent systems
(hence $p_{ij}^{A+B} = p_i^A p_j^B$ for every pair~$i,j$, consequently $W^{A+B}=W^{A}\,W^{B}$), 
we obtain ($k_B=1$ henceforth):
\begin{equation}
S_{BG}\de{A+B} =
- \sum_{i,j=1}^{W} p^{A+B}_{ij} \ln p^{A+B}_{ij} = S_{BG}\de{A} + S_{BG}\de{B}
\label{eq:SBG_Additive}
\end{equation}

To restrict ourselves only to systems of the exponential class appears as a rather
limiting and generically unjustified assumption. Indeed, strong correlations between the $N$ elements
do exist in a great variety of natural, artificial and social systems.
One can have, for example, systems belonging to the so called \emph{power-law class},
with the number of admissible microstates increasing like $W\de{N}\sim N^{\rho}\,\de{\rho>0}$. 
For this class, the additive entropy $S_{BG}$ is clearly not extensive, since it is proportional to $\ln N$.
Notice that $N^{\rho} \ll \mu^{N}$ for large $N$, 
which is intuitive since correlations tend to bind the system to a smaller number of accessible microstates.
In order to ensure extensivity for this kind of systems we shall use instead
the generalization~\cite{Tsallis1988, GellMannTsallis2004, TsallisBook2009}
of the BG entropy given by ($q \in \mathcal{R}$):
\begin{equation}
S_q = \sum_{i=1}^{W} p_i \ln_q \frac{1}{p_i} = -\sum_{i=1}^{W} p_i^q \ln_q p_i = -\sum_{i=1}^{W} p_i \ln_{2-q} p_i
\label{eq:Sq}
\end{equation}
In the $q \to 1$ limit, $S_q$ recovers $S_{BG}$ as seen in Eq.~\eqref{eq:SBG}; 
$\ln_q x  \equiv (x^{1-q} - 1)/(1 - q)$  (with $\ln_1 x = \ln x$) is the $q$-generalized logarithm. 

A remarkable property of $S_q$ is that it can be made extensive  for the power-law class with a suitable choice of the parameter~$q$.
If we look at its extremum value, also occurring when the probabilities are equal, \emph{i.e.}, when $p_i=1/W\de{N}$, $\forall\, i$,  
which leads to $S_q = \ln_q W\de{N}$, it is straightforward to verify that, if $W\de{N}\sim N^{\rho}$, $S_q \sim N$ as long as $q=1-1/\rho$,
result that can \emph{not} be achieved with~$S_{BG}$.

Most entropic functionals different from the Boltzmann-Gibbs one are nonadditive.
But it is precisely this nonadditivity which generically enables the entropy of the system to be extensive. There is in the literature a bit of confusion at
this respect\footnote{The confusion arises from the fact that, occasionally, some
authors inadvertently use nonadditive entropies for systems for which the entropy
to be used evidently is the BG one.
In the words of Tirnakli ``\emph{It is like trying to play golf with a soccer ball, and then complaining that it does not fit in the holes}" \cite{TirnakliPrivate2014}.} (see, for instance, \cite{PressePRL2013, TsallisArXiv2014}).
Equation~\eqref{eq:SBG_Additive} expresses the additivity of $S_{BG}$,
whereas the nonadditivity of $S_q$ is seen, as we may readily check, in
\[
S_q\de{A+B}
= S_q\de{A} + S_q\de{B} + \de{1-q}S_q\de{A} S_q\de{B}
\]
Inspired in the above result the so-called $q$-algebra~\cite{BorgesPA2004, Nivanen_etal2003} emerged.
In particular, for equal probabilities, we verify the additive-like property  $S_q\de{W^A \otimes_q W^B} = S_q\de{W^A}+S_q\de{W^B}$,
where the $q$-product $\otimes_q$ is defined in such way that $\ln_q\de{x\otimes_q y}= \ln_q x + \ln_q y$.

Let us now focus on  the \emph{stretched-exponential class},
another example of systems with strong correlations between its $N$ elements,  
where the number of admissible microscopic configurations increases like $W\de{N}\sim \nu^{N^\gamma}\, \de{\nu>1;  0<\gamma<1}$.
Therefore, the phase space is more restricted than the exponential class albeit being less restricted
than the power-law one, \emph{i.e.}, $N^{\rho}\ll \nu^{N^\gamma} \ll \mu^{N}$ for large $N$.
The (nonadditive) entropy which is able to provide extensivity for this class 
is\footnote{This
entropy was first proposed in~\cite{TsallisBook2009} (footnote on page 69) in order to construct an extensive entropy for the stretched-exponential class, and has been discussed in detail in~\cite{TsallisCirtoEPJC2013}.
The same functional form was also discovered (independently) by
Ubriaco in~\cite{UbriacoPLA2009}.} ($\delta>0$):
\begin{equation}
S_{\delta} = \sum_{i=1}^{W} p_i \de{\ln \frac{1}{p_i} }^{\delta}
\label{eq:S_Delta}
\end{equation}
Once again, the extremum of this entropic functional 
occurs for equal probabilities and it is straightforward to evaluate
that $S_\delta = [\ln W\de{N}]^\delta$ in this case.
For the specific value of~$\delta = 1/\gamma$ we verify that
$S_\delta \sim N$, hence extensive, a property which is unattainable
with $S_{BG}$ or $S_q$ for correlations within the stretched-exponential class.

We may now unify $S_q$ and $S_\delta$ to form
a new two-parameter entropic functional, namely, $S_{q,\delta}= \sum p_i\comu{\ln_q \de{1/p_i}}^{\delta}$
with $S_{1,\delta}=S_\delta$, $S_{q,1}=S_q$ and $S_{1,1}=S_{BG}$,
but we will not enter into details about this point here (see \cite{TsallisCirtoEPJC2013, RibeiroNobreTsallisPRE2014}). 
However it is worth to mention that $S_{q,\delta}$ can be also discussed within the framework of two-parameter entropies
advanced by Hanel and Thurner in~\cite{HanelThurnerEPL2011_1, HanelThurnerEPL2011_2}; see also Tempesta~\cite{TempestaPRE2011}.

The three classes of correlations discussed above, together with
their respective mathematical entropic functionals which yield an extensive entropy, 
are summarized in Table~\ref{Tab:Married_Boltzmann}. 
\begin{table}[h!] 
\centering 
\caption{Entropic functionals and classes of systems (\emph{exponential}, \emph{power-law} and \emph{stretched-exponential}, see text) 
for which the entropy is extensive, \emph{i.e.}, proportional to the number $N$ of elements.
$W(N)$ is the number of admissible microscopic configurations of a system with $N$ elements
(only configurations with nonvanishing occurrence probability are considered admissible).
We also see the specific values of~$q$ and $\delta$ for which respectively $S_q$ and $S_\delta$
are extensive.}
\begin{tabular}{|c|c|c|c|}   \cline{2-4} 
\multicolumn{1}{c|}{}                                &\multicolumn{3}{c|}{\rule[-3mm]{0mm}{9mm}\footnotesize{\textbf{ENTROPY}} } \\  \hline
\rule[-1mm]{0mm}{7mm}$\bm{W\de{N}}$                  &                   $S_{BG}$                                 & $S_{q}$                                                       & $S_{\delta}$                                                 \\
\rule[-2mm]{0mm}{7mm}$\bm{\de{N \to \infty}}$        &                                                            & $\de{q\neq1}$                                                 & $\de{\delta\neq1}$                                           \\
\rule[-3mm]{0mm}{7mm}                                & \textcolor{dark-green}{\textbf{\footnotesize{(ADDITIVE)}}} & \textbf{\textcolor{dark-green}{\footnotesize{(NONADDITIVE)}}} &\textbf{\textcolor{dark-green}{\footnotesize{(NONADDITIVE)}}} \\ \hline 
\Espaco$\displaystyle{{\sim\mu^{N}} \atop \de{\mu > 1} }$  &\textbf{\textcolor{blue}{\footnotesize{EXTENSIVE}}}   & \textbf{\textcolor{red}{\footnotesize{NONEXTENSIVE}}}  &\textbf{\textcolor{red}{\footnotesize{NONEXTENSIVE}}}              \\ \hline
\Espaco$\displaystyle{\sim N^{\rho} \atop \de{\rho > 0} }$ &\textbf{\textcolor{red}{\footnotesize{NONEXTENSIVE}}} & \textbf{\textcolor{blue}{\footnotesize{EXTENSIVE}}}    &\textbf{\textcolor{red}{\footnotesize{NONEXTENSIVE}}}              \\
\rule[-2mm]{0mm}{0mm}                                      &                                                      & $\de{q = 1- 1/\rho}$                                   &                                                                   \\ \hline
\Espaco$\displaystyle{\sim\nu^{ N^\gamma } \atop (\nu>1; }$&\textbf{\textcolor{red}{\footnotesize{NONEXTENSIVE}}} & \textbf{\textcolor{red}{\footnotesize{NONEXTENSIVE}}}  & \textbf{\textcolor{blue}{\footnotesize{EXTENSIVE}}}               \\
\rule[-2mm]{0mm}{0mm}$0<\gamma<1)$                         &                                                      &                                                        & $\de{\delta = 1/\gamma}$                                          \\ \hline
\end{tabular}

\label{Tab:Married_Boltzmann} 
\end{table}
\section{Why should the thermodynamical entropy always be extensive?}
\label{Sec:Thermodynamic_Entropy}
In this Section we will briefly summarize, along lines similar to those  of~\cite{TsallisFractals1995, TsallisCirtoEPJC2013, AbeRajagopalPLA2005}, the thermodynamic foundations underling the notion
that the entropy must always be an extensive quantity.
Let us first write a general Legendre-transformation form of a thermodynamical energy $G$ of a generic $d$-dimensional system ($d$ being an integer or fractal dimension):
\begin{eqnarray}
\begin{split}
G\de{V,T,p,\mu,H, \dots} & \,=\,  U\de{V,T,p,\mu,H,\dots} - TS\de{V,T,p,\mu,H,\dots} \,+ \\ 
  + &\,\, pV -\mu N\de{V,T,p,\mu,H,\dots} - HM\de{V,T,p,\mu,H,\dots} \,-\,\cdots
\label{eq:thermodynamics}
\end{split}
\end{eqnarray}
where $T, p, \mu,H$ are the temperature, pressure, chemical potential, external magnetic field,
and $U,S,V,N,M$ are the internal energy, entropy, volume, number of particles, magnetization.
We may identify three types of variables, namely 
(i) those that are expected to always be extensive ($S,V, N,M,\ldots$), \emph{i.e.}, scaling
with $V \propto L^d$, where $L$ is a characteristic linear dimension of the system
(notice the presence of~$N$ itself within this class), 
(ii) those that characterize the external conditions under which the system is placed ($T,p,\mu,H,\ldots$), scaling with $L^\theta$,
and (iii) those that represent energies ($G,U$), scaling with $L^\epsilon$.
Ordinary thermodynamical systems are those with~$\theta=0$ and $\epsilon=d$,
therefore both the energies and the generically extensive variables scale with~$L^d$ and
there is no difference between the types (i) and (iii) variables, being all of them extensive in this case.
There are, however, physical systems where~$\epsilon = \theta + d$ with~$\theta \neq 0$.
Let us divide Eq.~\eqref{eq:thermodynamics} by $L^{\theta + d}$, namely:
\begin{equation} 
\frac{G}{L^{\theta+d}} = \frac{U}{L^{\theta+d}}  - \frac{T}{L^\theta}\, \frac{S}{L^d} + \frac{p}{L^\theta} \frac{V}{L^d} - \frac{\mu}{L^\theta} \,\frac{N}{L^d}
-\frac{H}{L^\theta}\,\frac{M}{L^{d}} \,-\, \cdots
\label{eq:thermodynamics2}
\end{equation} 
If we consider now the thermodynamical $L\to\infty$ limit, we obtain
\begin{equation}
\widetilde{g} = \widetilde{u}  - \widetilde{T} s + \widetilde{p} v- \widetilde{\mu}\, n
-\widetilde{H} m \,-\, \cdots
\label{eq:thermodynamics3}
\end{equation}
where, using a compact notation, $\de{\widetilde{g}, \widetilde{u}} \equiv \lim_{L\to\infty} \de{G, U}/L^{\theta + d}$ represent the energies;
$\de{s, v, n, m} \equiv \lim_{L\to\infty} \de{S,V,N,M} /L^d$ represent the usual extensive variables
and $(\widetilde{T}, \widetilde{p}, \widetilde{\mu}, \widetilde{H}) \equiv \lim_{L\to\infty} \de{T, p, \mu, H}/L^{\theta}$
correspond to the usually intensive ones.
For a standard thermodynamical system (\emph{e.g.}, a real gas ruled by a Lennard-Jones short-ranged potential, a simple metal, etc)
we have $\theta =0$ (hence $(\widetilde{T},\widetilde{p},\widetilde{\mu},\widetilde{H}) = \de{T,p,\mu,H}$, \emph{i.e.}, the usual intensive variables),
and $\epsilon = d$ (hence $\de{\widetilde{g}, \widetilde{u}} = \de{g,u}$, \emph{i.e.}, the usual extensive variables);
this is of course the case found in the textbooks of thermodynamics.
Not yet really explored in textbooks are those cases with~$\theta \neq 0$.
Indeed, the correctness of the scaling appearing in Eq.~\eqref{eq:thermodynamics3} 
for nonstandard systems, \emph{i.e.}, for those with $\theta\neq0$, has been profusely 
verified for several systems in the literature~\cite{JundKimTsallisPRB1995, CannasTamaritPRB1996, SampaioAlbuquerqueFortunato1997PRB1997, AnteneodoTsallisPRL1998, TamaritAnteneodoPRL2000, GiansantiMoroniCampa2002CSF, Binek_etalPRB2006, thermal_5, thermal_6, thermal_7, CirtoAssisTsallisPA2014, CirtoLimaNobreArXiv2014, diffusion, geometrical, geometrical_2};
one of them is going to be discussed in Section~\ref{Sec:Classical_Spins} below.
Furthermore, it has been shown that such scalings preserve 
important thermodynamical relations such as the Euler and Gibbs-Duhem~\cite{AbeRajagopalPLA2005}.

The thermodynamic relations~\eqref{eq:thermodynamics} and \eqref{eq:thermodynamics2} put on an equal footing the entropy~$S$,
the volume~$V$ and the number of elements~$N$, and there can be no doubt about the extensivity of the latter two variables.
In fact, similar analysis can be performed using~$N$  instead of~$V$ since
$V\propto N$.
\begin{figure}[h!]
\centering
 \includegraphics[width=0.80\linewidth]{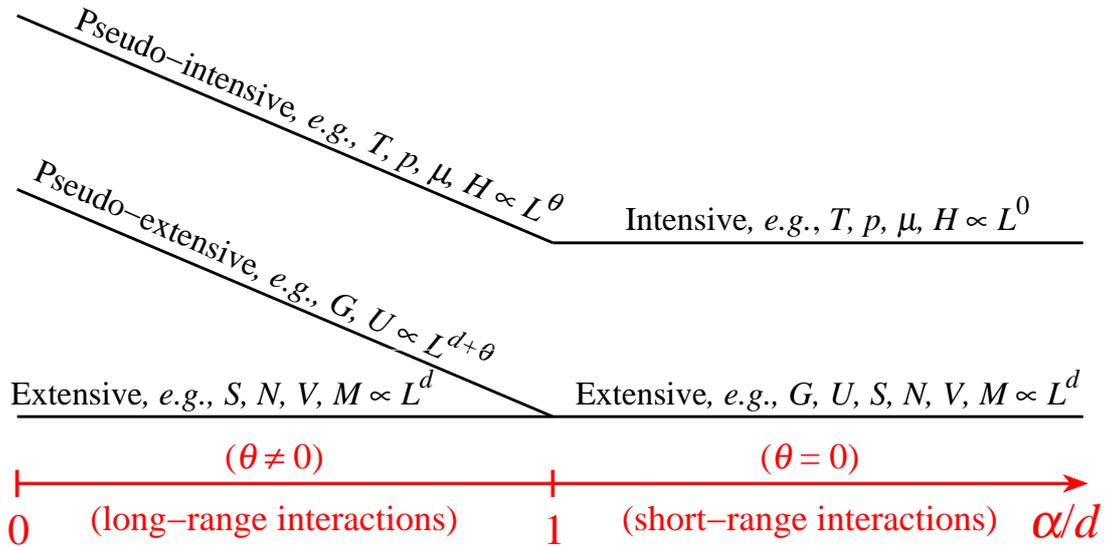}
 \vspace{0.3cm}
\caption{Representation of the different scaling regimes of the Eq.~\eqref{eq:thermodynamics2} for classical $d$-dimensional systems.
For attractive long-range interactions (\emph{i.e.}, $0 \leq \alpha/d \leq 1$, $\alpha$ characterizes the interaction range in a potential with the form $1/r^{\alpha}$)
we may distinguish three classes of thermodynamic variables, namely,
those scaling with $L^{\theta}$, named \emph{pseudo-intensive} ($L$ is a characteristic linear length, $\theta$ is a system-dependent parameter),
those scaling with $L^{d+\theta}$, the \emph{pseudo-extensive} ones (the energies),
and those scaling with $L^{d}$ (which are always extensive).
For short-range interactions (\emph{i.e.}, $\alpha > d$)
we have $\theta=0$ and the energies recover their standard~$L^{d}$ extensive scaling,
falling in the same class of~$S$, $N$, $V$, etc, 
whereas the previous pseudo-intensive variables become truly intensive ones (independent of $L$);
this is the region, with two classes of variables, that is covered by the traditional textbooks of thermodynamics.
}
\label{Fig:Pseudo}
\end{figure}

An example of nonstandard system with~$\theta\neq 0$ is the classical Hamiltonian discussed in Section~\ref{Sec:Classical_Spins} below. 
We consider two-body interactions decaying with distance $r$ like $1/r^\alpha \;(\alpha \ge 0 )$.
For this system we have~$\theta =d-\alpha$ whenever $0 \le \alpha < d$ (see, for example, Fig.~1 of~ \cite{TamaritAnteneodoPRL2000}).
This peculiar scaling occurs because the potential is not integrable,
\emph{i.e.}, the integral $\int_{\textrm{constant}}^\infty dr \,r^{d-1}\,r^{-\alpha}$
diverges for~$0 \le \alpha \le d$, therefore the Boltzmann-Gibbs canonical partition function itself diverges.
Gibbs was aware of this kind of problem and has pointed out~\cite{Gibbs1902} that whenever the partition function diverges, 
the BG theory can not be used because, in his words, ``the law of distribution becomes illusory".
The divergence of the total potential energy occurs for~$\alpha \le d$, which is referred to as long-range interactions. 
If $\alpha > d$, which is the case of the $d=3$ Lennard-Jones potential, whose attractive part corresponds to $\alpha=6$, 
the integral does not diverge and we recover the standard
behaviour of short-range-interacting systems with the~$\theta=0$ scaling.
Nevertheless, it is worth recalling that nonstandard thermodynamical behaviour is not necessarily associated with long-range interactions
in the classical sense just discussed.
A meaningful description would then be long-range correlations (spatial or temporal)
because for strongly quantum-entangled  systems, correlations are not necessarily connected with the interaction range (see Section~\ref{Sec:Quantum_Chain}).
However the picture of long- versus short-ranged interactions in the classical sense, directly related to the distance~$r$,
has the advantage to depict clearly the thermodynamic relations~\eqref{eq:thermodynamics} and ~\eqref{eq:thermodynamics2} for the different scaling regimes, as shown in Fig.~\ref{Fig:Pseudo}.

One more recent result is now
available~\cite{RuizTsallisPLA2012, TouchettePLA2013Comment, RuizTsallisPLA2013Reply},
related to the so called Large Deviation Theory in theory of probabilities~\cite{EllisBook1985,TouchettePR2009}, which also is 
consistent with the extensivity of the entropy, even in the presence of strong correlations
between the elements of the system.
In fact it is known since several
decades that the mathematical foundation of BG statistical mechanics 
crucially lies on the theory of large deviations. 
To attain the same status for nonextensive
statistical mechanics, it is necessary to $q$-generalize the large deviation 
theory itself. The purpose of those efforts precisely is to
make a first step towards that goal through the study of a simple model.

Finally, a further indication we can refer to is the analogy with the time $t$ dependence
of the entropy of simple nonlinear dynamical systems, \emph{e.g.}, the logistic map.
Indeed, for the parameter values for which the system has positive 
Lyapunov exponent (\emph{i.e.}, strong chaos and ergodicity), 
we verify~$S_{BG}~\propto~t$ (under appropriate mathematical limits), 
but for parameter values where the Lyapunov exponent vanishes nontrivially, \emph{e.g.}, the Feigenbaum point
(\emph{i.e.}, weak chaos), it is the nonadditive entropy $S_q$ for a specific value of $q$ the one which grows linearly with $t$ 
(see~\cite{BaldovinRobledo2004, BaldovinRobledo2004_2, BaldovinRobledo2004_3, BaldovinRobledo2004_4, BaldovinRobledo2004_5, BaldovinRobledo2004_6, BaldovinRobledo2004_7, BaldovinRobledo2004_8, BaldovinRobledo2004_9, BaldovinRobledo2004_10}
and references therein),
and consistently provides a generalized Pesin-like identity.
If we take into account that, in many such dynamical systems, 
$t$ plays a role analogous to $N$ 
in thermodynamical systems, we have here one more indication which 
aligns with the extensivity of the entropy for complex systems.

In what follows we illustrate the above concepts through three physical systems, namely a long-range-interacting many-body classical Hamiltonian system (Section \ref{Sec:Classical_Spins}), a strongly quantum entangled system at zero temperature (Section \ref{Sec:Quantum_Chain}), and black holes (Section \ref{Sec:blackhole}).
\section{A classical model with long- and short- ranged interactions} 
\label{Sec:Classical_Spins}
To better discuss the concepts of long- and short-range interaction let us see a concrete
and well known example, namely, an ensemble of~$N$ classical spins arranged in a lattice
whose Hamiltonian is given by
\begin{equation}
\mathcal{H} = -J \sum_{\left\langle i,j \right\rangle}^N \mathbf{S}_i\cdot\mathbf{S}_j
\label{eq:Ising_Hamiltonian}
\end{equation}
where~$J>0$ is the ferromagnetic coupling constant.
The symbol $\left\langle i,j \right\rangle$ means that the sum runs only over the 
nearest neighbour for each spin. 
If the system lies in a ring, \emph{i.e.}, a one-dimensional system, each spin 
has only two nearest neighbours; if it lies in say bidimensional plane and is arranged
as a square lattice there are four nearest neighbours. 
It is a typical example of what is referred to as \emph{short}-range interactions.
Depending on the dimension of the spin vector $\mathbf{S}_i$, 
the Hamiltonian~\eqref{eq:Ising_Hamiltonian} may represent
the Ising, the classical~XY or Heisenberg models,
all of them very well understood and described by the traditional
Boltzmann-Gibbs equilibrium statistical mechanics (see, for example, the classical paper by Stanley~\cite{StanleyPR1969} for the linear case).

Let us now consider the case where the interaction is not restricted to the
nearest neighbours anymore. Let us consider the other extreme situation where all spins
interact with all others with the same strength regardless the distance between them.
The system is said to be fully-coupled and is described by the following Hamiltonian:
\begin{equation}
\mathcal{H} = -J \sum_{i=1}^N\sum_{j=1}^N \mathbf{S}_i\cdot\mathbf{S}_j
\label{eq:Fully_Coupled_Hamiltonian}
\end{equation}
It is a typical example of a long-range system.
It may be directly assessed that the total energy of this model
is not proportional to the system size~$N$, hence the system is nonextensive
and, strictly speaking, there is no traditional thermostatistics in this case at all.
In the framework of Section~\ref{Sec:Thermodynamic_Entropy}, 
Hamiltonian~\eqref{eq:Fully_Coupled_Hamiltonian} is associated with~$\theta=d$
which means a $U\propto V^{2d}\propto N^{2}$ scaling.
Accordingly, there is no quantity different from zero or from infinity that can be calculated
(``the law of distribution becomes illusory" here).

A way to overcome the diff{}iculty without moving out of the standard formalism can be thought of.
If the (initially) constant coupling constant is conveniently rescaled as $J \to J/N$
--- nowadays called Kac's prescription ---, the extensivity of the system is recovered.
Mathematically this procedure is evidently rightful but it 
throws us in an strange situation where the microscopic coupling constant~$J$ become
dependent on~$N$, \emph{i.e.}, following Baxter's words~\cite{BaxterBook}, it leave us with the ``unphysical property that the interaction strength depends on the number of particles''.

Leaving aside the epistemological issue of having a two-body coupling constant~$J$
dependent on the system~$N$ let us put forward the scaling~$J \to J/N$ idea
and explore the following system~\cite{AntoniRuffo1995,AnteneodoTsallisPRL1998, TamaritAnteneodoPRL2000, CampaGiansantiMoroni2000PRE, CirtoAssisTsallisPA2014}:
\begin{equation}
\mathcal{H}=\frac{1}{2}\sum_{i=1}^N p_i^2 + \frac{J}{2\Ntil}\sum_{i = 1}^N \sum_{ \substack{j=1 \\ j \neq i} }^N \frac{1- \cos\de{\theta_i - \theta_j} }{r_{ij}^\alpha}
\label{eq:Hamiltonian_Alpha_XY}
\end{equation}
The Hamiltonian~\eqref{eq:Hamiltonian_Alpha_XY} is a extension of the 
models~\eqref{eq:Ising_Hamiltonian} and~\eqref{eq:Fully_Coupled_Hamiltonian}
by including a kinetic term provided that the classical spin vectors~$\mathbf{S}_i$ be bidimensional, since, in this case, $\mathbf{S}_i\cdot\mathbf{S}_j = \cos\de{\theta_i-\theta_j}$.
In the $\alpha\to\infty$ limit, the interaction term of Eq.~\eqref{eq:Hamiltonian_Alpha_XY} approaches the first-neighbours Hamiltonian~\eqref{eq:Ising_Hamiltonian},
whereas if $\alpha = 0$ it approaches the fully-coupled instance~\eqref{eq:Fully_Coupled_Hamiltonian}.
The latter case is called Hamiltonian Mean Field, or simply HMF, after Antoni and Ruffo' s work~\cite{AntoniRuffo1995},
a model that has been profusely studied in the past years.

The parameter $\Ntil$ is judiciously chosen in order to make the energy extensive for all values of $\alpha/d$.
It behaves, when $N$ is large, like $\Ntil\sim N^{1-\alpha/d}$ if  $0 \le \alpha/d <1$
and like $\Ntil\sim\mathcal{O}\de{1}$ if $\alpha/d>1$.
In other words, it recovers the general scaling for systems with~$\alpha \neq 0$.
This is referred to as the $\alpha$-XY model (the spins may be visualized as XY-planar rotators), and it is a genuine Hamiltonian system
in the sense that the variables~$p_i$ and $\theta_i$ are canonical conjugate pairs.
With a kinetic term the model presents its own dynamics and
equation of motion can be derived throughout a Hamiltonian formulation. Consequently, to enquire numerically physical properties of the system through molecular dynamic simulations constitutes a natural route.

\begin{figure}[h!]
\centering
\includegraphics[width=0.495\linewidth]{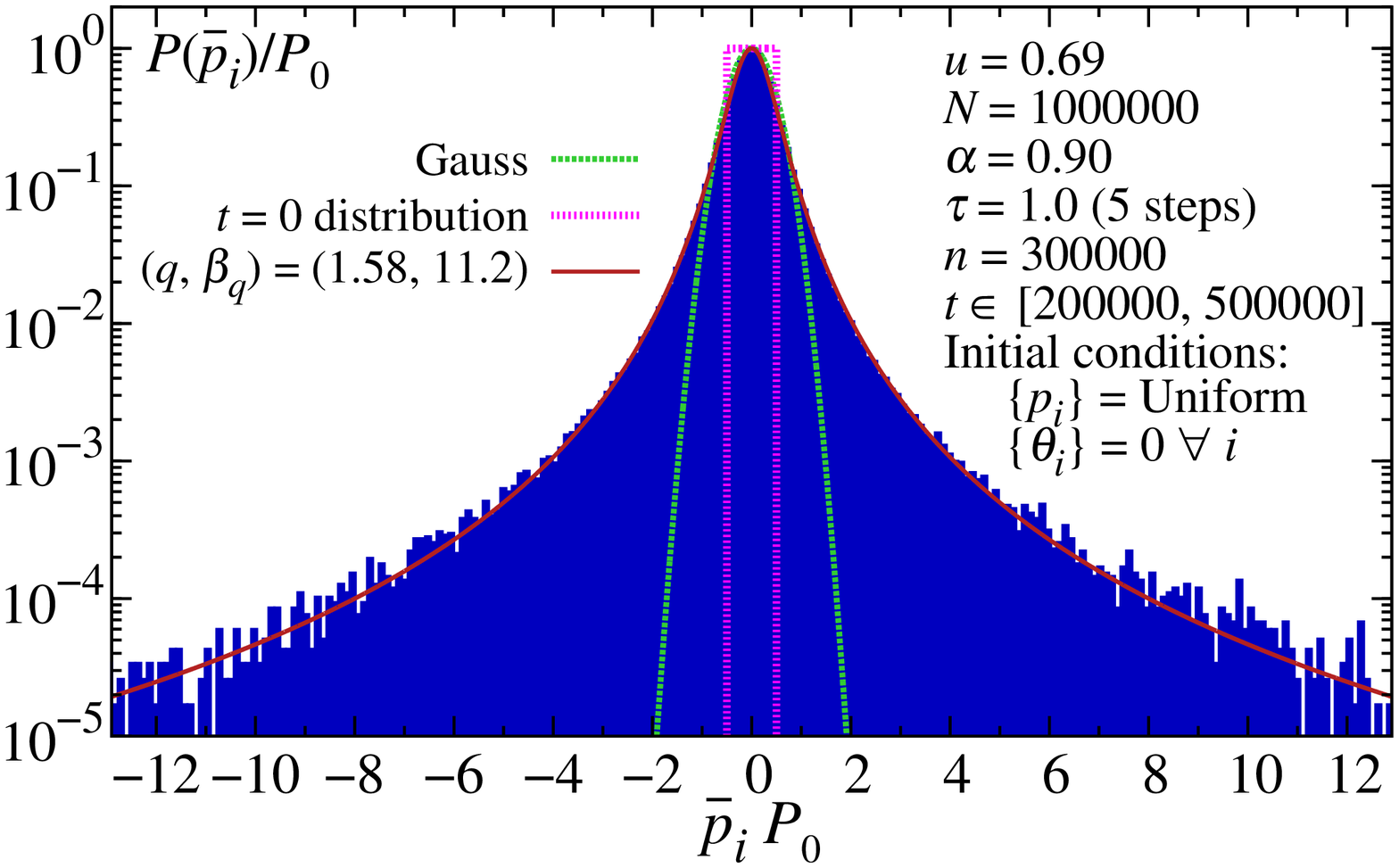}
\includegraphics[width=0.495\linewidth]{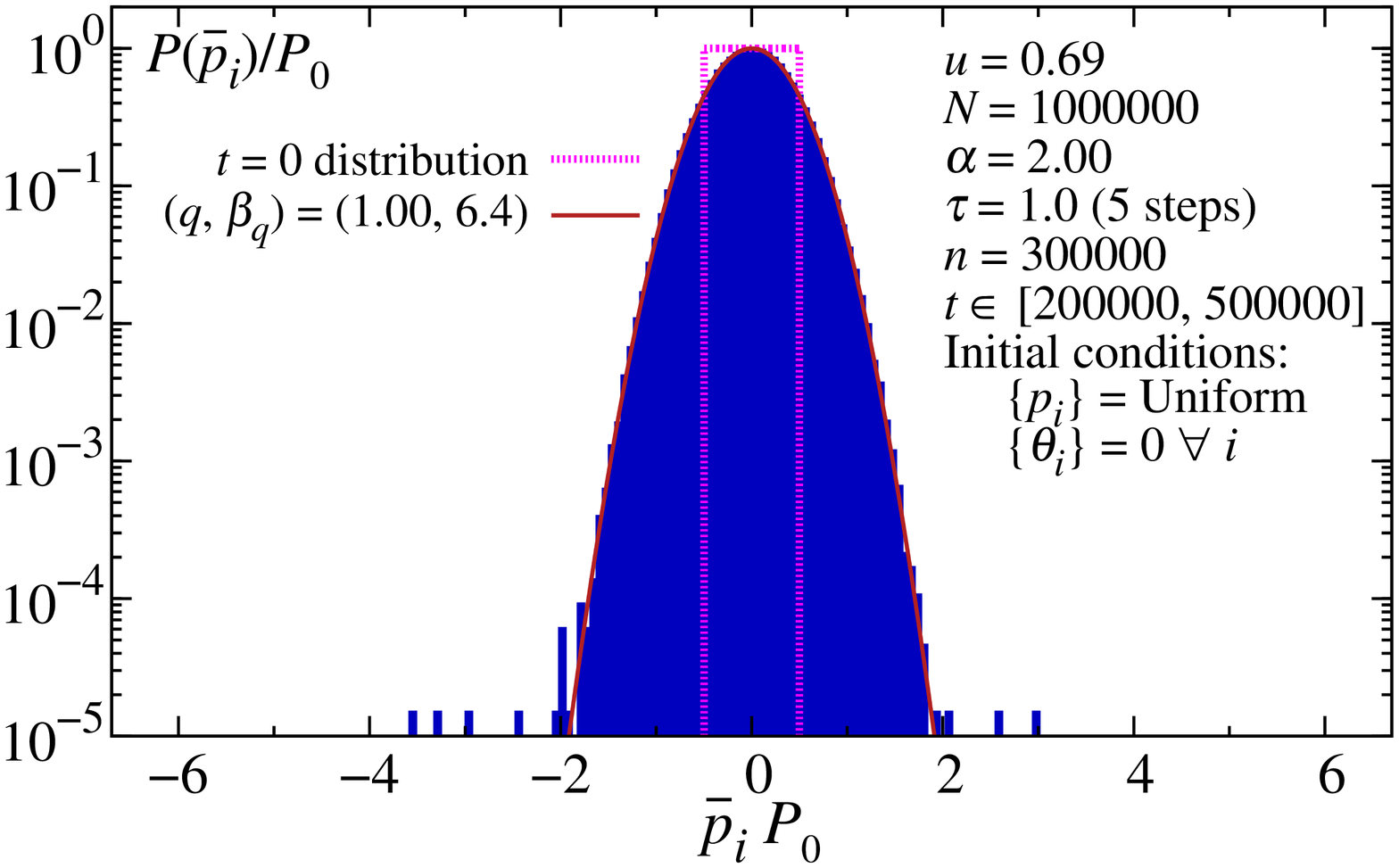}
\caption{Molecular dynamics results for the $d=1$ Hamiltonian~\eqref{eq:Hamiltonian_Alpha_XY} for two values of $\alpha$ under identical
simulational setup for all the other parameters (energy, initial conditions, number of particles, period over which the time average is calculated, etc; we consider here, without loss of generality, $J=1$).
What we see is a typical single-initial-condition one-momentum distribution for $\alpha=0.9<d=1$ (long-range, left plot) and $\alpha=2.0 >d=1$ (short-range, right plot).
The continuous curves correspond to $q$-Gaussians 
with $q = 1.58$ for $\alpha=0.9$ and $q=1$ (\emph{i.e.}, a Gaussian) for $\alpha = 2.0$.
See details in~\cite{CirtoAssisTsallisPA2014}.
}
\label{Fig:q_Gaussiana}
\end{figure}

After scaling the interaction with~$\Ntil$, 
the energy of the system becomes extensive and all the traditional thermodynamical techniques 
can be applied (the canonical partition function does not diverge in the thermodynamic limit anymore).
It remains, however, nonadditive, \emph{i.e.},
if we bring together two system~$A$ and $B$ ruled by Hamiltonian~\eqref{eq:Hamiltonian_Alpha_XY},
the joined internal energy~$u_{A+B}$, with $u = \media{\mathcal{H}/N}$, will be 
$u_{A+B} \neq u_A + u_B$ in general.
This happens because the long-range nature of the interaction is still present,
property particularly seen when $\alpha=0$ and the scaling decreases the interaction strength equally regardless the distance.
Thereby, even with the extensivity recovered, 
unexpected behaviour should not be seen as striking, and
it has been actually caught in several (numerical) experiments.
As one example of unpredict behaviour within the traditional scenario, one has 
the long-lived quasistationary states
(QSS) which emerges for $\alpha/d < 1$~\cite{LatoraRapisardaRuffoPRL1998, LatoraRapisardaRuffoPD1999, LatoraRapisardaTsallisPRE2001, CampaGiansantiMoroniPA2002, PluchinoLatoraRapisardaPA2004, PluchinoLatoraRapisardaPD2004, PluchinoLatoraRapisardaPA2006, Yamaguchi_etalPA2004, MoyanoAnteneodoPRE2006, PluchinoRapisardaTsallisPA2008, EttoumiFirpoPRE2013, CirtoAssisTsallisPA2014}.
In these states the thermodynamical quantities like temperature and magnetization
do not coincide with the canonical predictions.
Moreover, its lifetime diverge with increasing system size~$N$, associated with 
the order in which the thermodynamic ${N\to\infty}$ and the infinite time ${t\to\infty}$ limits are
considered. Specifically, if we let ${N\to\infty}$ first, the system remains trapped in these
QSS's, never reaching the final Boltzmann-Gibbs equilibrium state,
most probably being the QSS itself the ultimate state in this case.
Another example comes from the one-momentum distribution.
Within the BG framework it is expected a Maxwellian distribution for the velocities, no matter
whether it is calculated by using time or ensemble averages.
This distribution shape, \emph{i.e.}, a Gaussian, is in fact observed in the (time average) numerical simulations, but only for $\alpha$ suff{}iciently large (hence short-range).
However, if $\alpha$ is small (hence long-range), it was 
observed~\cite{PluchinoRapisardaTsallisEPL2007, PluchinoRapisardaTsallisPA2008, CirtoAssisTsallisPA2014}
distributions very well described by $q$-Gaussians, in disagreement with the
traditional BG thermostatistical scenario (see Fig.~\ref{Fig:q_Gaussiana}).

The model~\eqref{eq:Hamiltonian_Alpha_XY} is very rich and certainly will 
continue giving rise to several new and interesting results.
These nonstandard behaviours observed for $\alpha/d < 1$ appear to be 
in line with nonergodicity~\cite{EttoumiFirpoPRE2013, CampaDouxoisRuffoPR2009, ChavanisCampaEPJB2010, CampaChavanisEPJB2013}
and with the thesis of the $q$-generalized Central Limit Theorem (see references and comments in~\cite{CirtoAssisTsallisPA2014}).
The fact that the a \emph{ad hoc} scaling of the interaction recovers the formal extensivity
but not eliminate its intrinsic long-range nature
appears to be the reason why this model is not satisfactorily described within the BG thermodynamics scenario.
It is conceivable that such long-range interaction could 
generate correlations strong enough 
to constraint the dynamics of the system within some regions of the phase space,
thus reducing the ``number" of accessible microstates
in the same spirit of the correlation classes discussed in Section~\ref{Sec:Entropies}.
\subsection{Searching for \texorpdfstring{$q$}{q} from first principles}
It is expected that the index~$q$ of the entropy functional~$S_q$ shown in Eq.~\eqref{eq:Sq} to be an intrinsic property of the 
geometrical/dynamical nature of the occupancy of phase space.
It should be calculated from first principles, \emph{i.e.}, from the microscopic fundamental dynamical law governing  the system.
However this calculation is by no means an easy task and, in many cases, it will be virtually impossible without strong mathematical approximations.
Nevertheless a few examples have been analytically solved wherein a first-principle $q$ value was achieved,
as the one discussed in Section~\ref{Sec:Quantum_Chain} bellow.
Here, for the many-body $\alpha$-XY model, $q$ shall be approached through the Hamiltonian~\eqref{eq:Hamiltonian_Alpha_XY} itself.

Inspired by the $q$-Gaussian one-momentum distribution seen in Fig.~\ref{Fig:q_Gaussiana}
we may figure out a possible route to calculate from first principles the value of~$q$.
This distribution extremize the nonadditive entropy $S_q$ upon which the nonextensive statistical
mechanics~\cite{Tsallis1988, GellMannTsallis2004, TsallisBook2009} is based.
Within this framework, the stationary state is expected to yield a probability distribution 
$\exp_q\de{-\beta_q \mathcal{H} } / Z_q(\beta_q)$ with $Z_q(\beta_q)$
being the generalized partition function ($\exp_q$ represents the inverse of the $q$-generalized logarithm defined in Section~\ref{Sec:Entropies}; the $\exp_q$ function becomes the ordinary exponential for $q=1$).
The one-momentum marginal probability would then be calculated using
$P(p_1)= \int\!\rd p_2...\rd p_N\,\rd\theta_1...\rd\theta_N \exp_q\de{-\beta_q \mathcal{H}}/Z_q$.
The possible functional form of $P(p_1)$ could be 
a $q_{m}$-Gaussian, where $m$~stands for momentum
(we singled the label out here because the value $q_m$ is not necessarily the same as that of the entropic functional~$S_q$; naturally we expect $q_m=1$ if $q=1$).
The entropic index $q$ (and also $q_m$) is expected to characterize universality classes,
possibly a function $q=q(\alpha/d)$ to be different from $1$ for $0 \le \alpha/d <1$, and 
equal to~$1$ for $\alpha/d \ge 1$ in accordance with the numerical experiments.
At the present computational stage, we have access to~$q_m$ but not yet to~$q$.
The latter implies an extremely heavy computational task since it has to do with the occupancy of the entire many-body phase space for given initial conditions.
\section{A fully quantum-entangled system - An exact calculation of \texorpdfstring{$q$}{q} from first principles}
\label{Sec:Quantum_Chain}
When dealing with classical systems, spatial correlations and
long-time memory are usually neatly connected with long-range
interactions. However, when one goes deeper in the microscopic
structure of the matter the strictly
quantum mechanical phenomenon of entanglement comes into play and
long-range correlations are not necessarily connected with long-range
interactions in the sense discussed in the
Section~\ref{Sec:Classical_Spins}.
For example, let us consider the following first-neighbourhood
interaction Hamiltonian describing a quantum $N$ spin-$1/2$
ferromagnetic chain under a transverse magnetic field at its critical
value at zero temperature:
\begin{equation}
H = -\sum_{i=1}^{N-1} \comu{\de{1+\gamma}\sigma_i^x\sigma_{i+1}^x +
\de{1-\gamma}\sigma_j^y\sigma_{i+1}^y + 2\lambda\sigma_i^z}
\label{eq:Quantum_Spin_Chain}
\end{equation}
where~$\sigma^{\mu}$, $\mu=x,y,z$, are the Pauli's matrices, and
$\gamma$ and $\lambda$
are the intensity of the anisotropy and magnetic field respectively.
Known as quantum XY~model, Hamiltonian~\eqref{eq:Quantum_Spin_Chain}
recovers for $|\gamma| = 1$ (\emph{i.e.}, maximum axial anisotropy) the
quantum Ising chain.
Furthermore it is known that, in the thermodynamic~$N\to\infty$ limit,
a quantum phase transition (hence at $T=0$) exists at the critical
point $|\lambda_c| = 1$.

If we have complete information about a system its
entropy is zero.
Quantum mechanically, complete information means that we are dealing with
a pure state.
Evidently the intrinsic probabilistic nature of a quantum system
forbids us to have complete information in a classical sense;
a pure state means that there is an unique quantum \emph{state}
describing the system.
At zero temperature the system is in its fundamental state,
hence the entropy should be zero for any admissible entropic functional.
Being $\widehat{\rho}_N$ the density operator of the whole chain, a
pure state means
that~$\mathrm{Tr}\,\widehat{\rho}_N^{\,2} = \mathrm{Tr}\,\widehat{\rho}_N=1$.
However, even for $T=0$, it is possible to calculate a entropy
different from zero if we consider only a block of~$L$ contiguous spins
and work with the reduced matrix $\widehat{\rho}_L = \mathrm{Tr}_{N-L}
\widehat{\rho}_N$.
This reduced matrix in general does not represent a pure state, but
a mixed state instead (\emph{i.e.}, $\textrm{Tr}\,\widehat{\rho}_L^{\,2} < \textrm{Tr}\,\widehat{\rho}_L = 1$).
This fact is a consequence of the nonlocal entanglement that is
responsible  for
the long-range quantum correlations of the spin
chain~\eqref{eq:Quantum_Spin_Chain} at $T=0$.

\begin{figure}[h!]
\centering
  \includegraphics[width=0.60\linewidth]{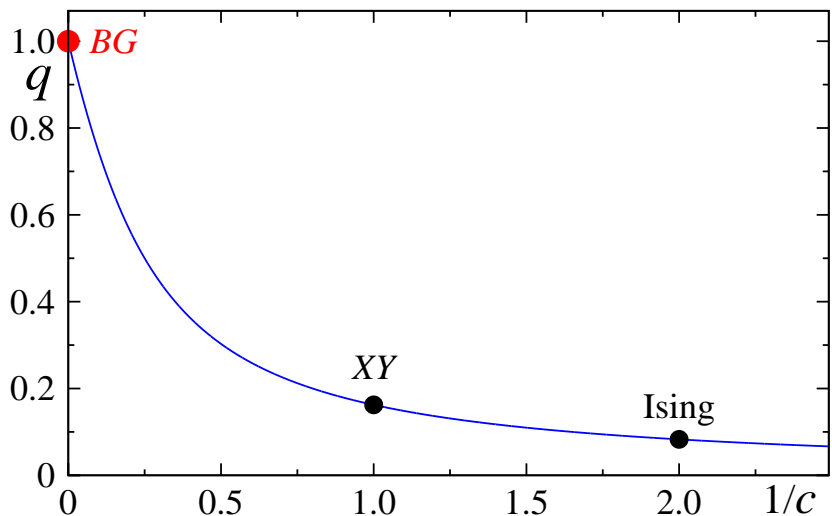}
\caption{$q$ as a function of the central charge~$c$ which characterizes the
universality class (and which contains the quantum spin
chain~\eqref{eq:Quantum_Spin_Chain} as a particular instance).
The BG entropy for a block of $L$ contiguous spins is $S_{BG}\de{L}
\propto \ln L$ for all finite values of the central charge, thus
violating thermodynamical extensivity. However, for the special values
of~$q$ shown here (see Eq.~\eqref{eq:q_Block_Entropy}), $S_q\de{L}
\propto L$, \emph{i.e.}, it satisfies one-dimensional extensivity,
thus enabling the use of all the relations that can be found in any
good textbook of thermodynamics.
See details in~\cite{CarusoTsallisPRE2008}.}
\label{Fig:Caruso}
\end{figure}

The degree of entanglement between a block of~$L$ contiguous spins and
the rest of the chain in its ground state
can be characterized by the von Neumann
entropy~\eqref{eq:Entropia_von_Neumann}
of the block (see~\cite{EntanglementRMF2008, EisertCramerPlenioRMP2010}).
For a large block size, it typically saturates off criticality, whereas
it is \emph{logarithmically} unbounded at the critical point,
\emph{i.e.}, the so called \emph{area law} for $d=1$ systems. In order
words, the BG entropy at $T=0$ for the
chain~\eqref{eq:Quantum_Spin_Chain} does not scale with the system size,
but like $S_{BG}\de{L}\propto \ln L$,
being $S_{BG}\de{L}\propto L^{d-1}$ the general area-law scaling for
$d$-dimensional systems with $d>1$.
However, it was shown~\cite{CarusoTsallisPRE2008} (see
also~\cite{SaguiSarandy2010}) that the thermodynamical extensivity is
recovered if we move from the BG entropic functional to the~$S_q$ one.
Furthermore, as the Hamiltonian~\eqref{eq:Quantum_Spin_Chain} can be
exactly diagonalized, is was possibly to calculate analytically, for the universality class
characterized by the central charge $c$, a closed form for~$q$, namely
\begin{equation}
q = \frac{\sqrt{9+c^2} - 3}{c}
\label{eq:q_Block_Entropy}
\end{equation}
Therefore, in order to achieve an extensive entropy for the
chain~\eqref{eq:Quantum_Spin_Chain},
which implies a \emph{finite} value for $S/L$ in the $L\to\infty$ limit,
it is enough to calculate $S_q$ with the specific value
of~$q$ shown in Eq.~\eqref{eq:q_Block_Entropy} (see Fig.~\ref{Fig:Caruso}).
\section{On the entropy for black holes}
\label{Sec:blackhole}
Far from going into details on this fascinating topic,
which very recently had its foundations expanded by one of its most
important
contributors~\cite{HawkingArXiv2014, MeraliNatureNews2014},
we will use the \emph{black hole} physical system  as a possible
application of the entropy~$S_{\delta}$ discussed in
Section~\ref{Sec:Entropies}~(Eq.~\eqref{eq:S_Delta}).
As already discussed, the Boltzmann-Gibbs entropy has, as underling
hypothesis, weak 
correlations and ergodicity.
To fit a black hole under this general assumptions may eventually not
be a safe thermodynamical starting
approach. Indeed, the outstanding results of Bekenstein and
Hawking~\cite{BekensteinPRD1973, HawkingPRD1976, BekensteinCP2004}
have already shown that the BG entropy of a black hole is proportional
to its boundary surface.
It is important to recall that, for a variety of reasons, this result
appears to be evidently true,
\emph{i.e.}, the BG entropy is in fact proportional do the area, as
several and diversified calculations along almost forty years have
confirmed~\cite{Solodukhin}.
In a few words, the Bekenstein-Hawking result reads:
\begin{equation}
S_{BH} \propto A
\label{eq:SBH}
\end{equation}
where $A$ is the event horizon area.
\emph{If} the black hole is to be considered as a genuine~$d=2$ system,
which means that it is physically identified solely with its event
horizon surface, Eq.~\eqref{eq:SBH} is extensive and must be seen as the the truly
thermodynamical entropy.
Therefore, in a thermodynamical sense, there is nothing that should be
regarded as intriguing or unusual, and, strictly speaking,
this would not be an area-law
problem~\cite{EisertCramerPlenioRMP2010,DasShankaranarayananPRD2006, BrusteinEinhornYaromJHEP2006,KolekarPadmanabhanPRD2011}.
However, \emph{if} the black hole is to be considered a genuine~$d=3$
system, we then recover the very same discussion of the previous
Section~\ref{Sec:Quantum_Chain}.
Its thermodynamical entropy then should not be associated with the
additive BG functional and a nonadditive generalization
should be used instead.
It happens that the Bekenstein-Hawking result~\eqref{eq:SBH} is very
helpful here too, since it says to us that~$W \propto e^{bA}$ ($b>0$), hence we are dealing
with a stretched-exponential class system.
Therefore, it follows that extensivity is recovered by using the entropic
functional~$S_\delta$ with $\delta=3/2$ (see more details in~\cite{CirtoAssisTsallisPA2014}).

This idea has recently been put forward by Komatsu and
Kimura~\cite{KomatsuKimuraPRD2013,KomatsuKimuraPRD2014}
within a entropic-force scenario. See
also~\cite{RibeiroTsallisNobrePRE2013}, where the probability
distribution that extremizes $S_\delta$ and its associated
Fokker-Plank equation are analyzed.
\section{Final remarks}
To conclude, let us now summarize the line of thought that we have presented here.
Thermodynamics is a highly valuable approach to nature, and we see no reason at all
for generalizing its basic principles, in particular in what concerns entropy.
Because of the Legendre-transform structure of thermodynamics, as well as because
of strong indications within the probabilistic large-deviation theory, the extensivity
of the entropy must be preserved in all cases that we are aware of, whether short- or long-ranged-interacting
systems, classical or quantum (strongly entangled or not), dissipative systems, among others.
For systems that live in their entire (or nearly entire) phase space (or Hilbert or Fock spaces if the system is a quantum one),
in other words, if the system is essentially ergodic in a region with finite Lebesgue measure, the number~$W(N)$ of admissible microscopic
possibilities increases exponentially with~$N$ (exponential class).
Consequently it is the Boltzmann-Gibbs (\emph{additive}) entropic 
functional which guarantees the extensivity of the entropy.
But if, due to strong correlations between the elements of the system, the occupancy of phase space is severely restricted
(so strongly that the Lebesgue measure of the visited region is zero, which is the case of the power-law and stretched-exponential classes), we typically
need \emph{nonadditive} entropic functionals such as~$S_q$ or~$S_\delta$ in order to comply with the requirement of extensivity for the thermodynamical entropy.
This fact has very relevant consequences, in particular in what concerns the probability distributions that spontaneously emerge in the
corresponding (frequently unique) stationary or quasi-stationary states.
They tend to exhibit, for example for the power-law class, $q$-exponential behaviors (which asymptotically are power-laws)
instead of the usual exponential ones that typically are observed for all kinds of relevant physical properties.
A vast literature illustrates this scenario~\cite{bibliography}.
We have here selected a few of such examples (classical long-range-interacting many-body Hamiltonian systems, strongly
quantum entangled systems at their quantum critical point, black holes).
The same picture is seen in many other systems through analytical, experimental, observational and computational results
in natural, artificial and social systems, along with predictions, verifications and applications (see, for instance, a brief review in~\cite{Tsallis2014}).
Further reinforcing and clarifying evidences are naturally welcome.
Indeed, a variety of open questions (whose details are out from the present scope) still remain to be better understood.
\subsection*{Acknowledgments}
We have benefited from fruitful remarks by M.\ Jauregui and U.\ Tirnakli.
We also acknowledge partial financial support from CNPq, Faperj and Capes (Brazilian agencies).
It is our great pleasure to dedicate this review to \textbf{Hans J. Herrmann},
wishing him a very happy anniversary for his (first) 60 years!

\end{document}